\documentclass[preprint,12pt]{elsarticle}

\usepackage[utf8]{inputenc}
\usepackage{subfigure}
\usepackage{amsmath}
\usepackage{amssymb}
\usepackage{array}
\usepackage[version=3]{mhchem}
\usepackage{lineno}
\usepackage{pdfsync}

\newcommand{\care}{(Ca$_{1-x}RE_x$)$_{10}$(FeAs)$_{10}$(Pt$_3$As$_8$)}
\newcommand{\parent}{Ca$_{10}$(FeAs)$_{10}$(Pt$_3$As$_8$)}
\newcommand{\zva}{Ca$_{10}$(FeAs)$_{10}$(Pt$_4$As$_8$)}
\newcommand{\caeu}{(Ca$_{1-x}$Eu$_x$)$_{10}$(FeAs)$_{10}$(Pt$_3$As$_8$)}
\newcommand{\cala}{(Ca$_{1-x}$La$_x$)$_{10}$(FeAs)$_{10}$(Pt$_3$As$_8$)}

\journal{Solid State Communications}

\begin{document}
\begin{frontmatter}

\title{Superconductivity by rare earth doping in the 1038-type compounds  \care\ with $RE$ = Y, La-Nd, Sm-Lu}

\author{Tobias Stürzer}
\author{Gerald Derondeau}
\author{Dirk Johrendt\corref{cor1}}
\ead{johrendt@lmu.de}
\cortext[cor1]{Corresponding author}
\address{Department Chemie, Ludwig-Maximilians-Universität München, Butenandtstr. 5-13 (Haus D), 81377 München, Germany}


\begin{abstract}
We report superconductivity in polycrystalline samples of the 1038-type compounds \care\ up to $T_c$ = 35 K with $RE$ = Y, La - Nd, Sm, Gd - Lu. The critical temperatures are independent of the trivalent rare earth element used, yielding an universal $T_c(x_{RE})$ phase diagram for electron doping in all these systems. The absence of superconductivity in Eu$^{2+}$ doped samples, as well as the close resemblance of \care\ to the 1048 compound substantiate that the electron doping scenario in the $RE$-1038 and 1048 phases is completely analogous to  other iron-based superconductors with simpler crystal structures.
\end{abstract}


\end{frontmatter}

\section{Introduction}
\begin{figure}[h]
\centering
\includegraphics[width=7cm]{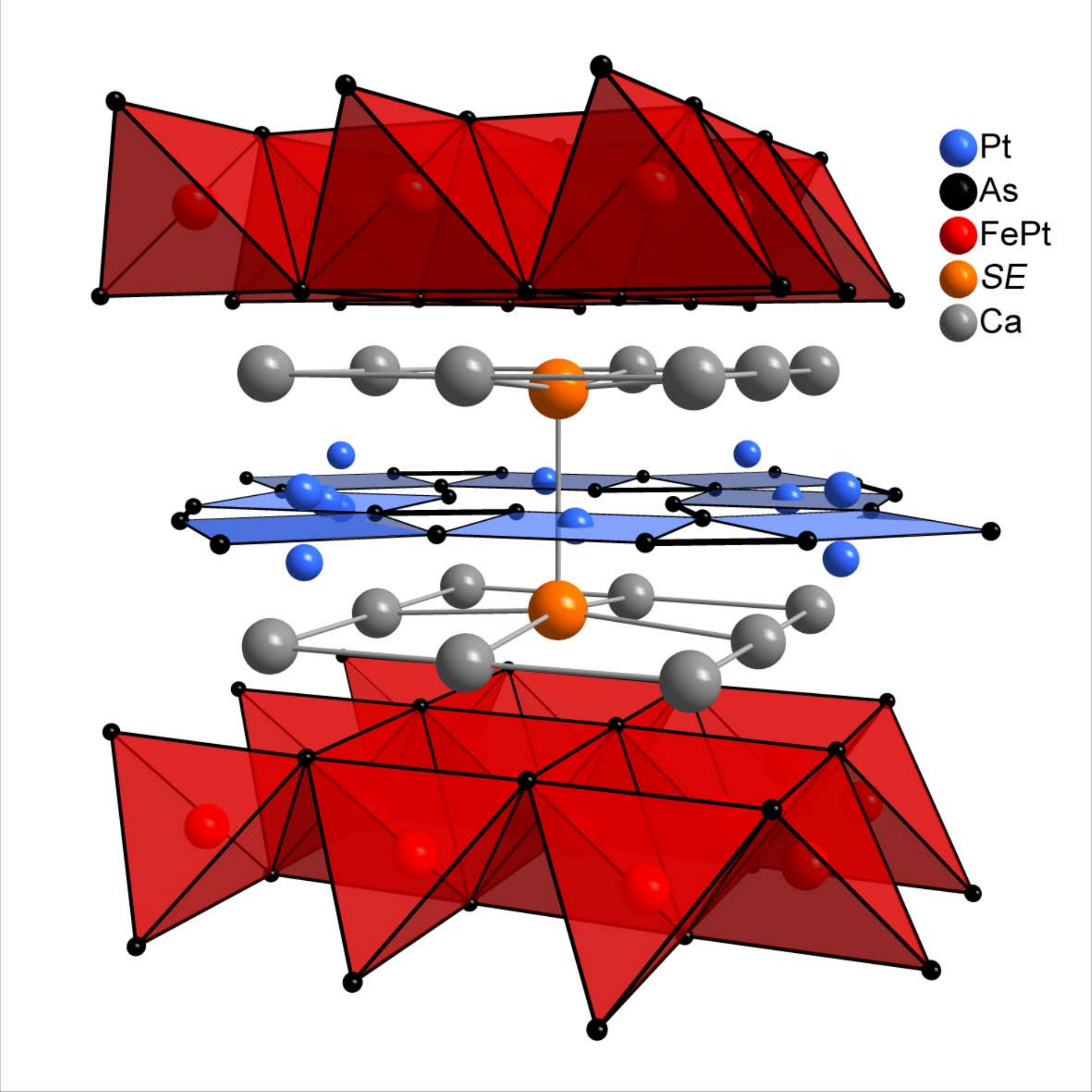}
\caption{\label{fig:structure} Crystal structure of \care.}
\end{figure}

Ever since the discovery of superconductivity in iron pnictides \cite{Kamihara-2008}, electronic doping is known to be an effective method to suppress the antiferromagnetic ground state of these compounds and induce high-$T_c$ superconductivity \cite{Johnston-2010,Stewart-2011,Johrendt-2011}. This was confirmed for many iron arsenides, examples are  La(O$_{1-x}$F$_x$)FeAs \cite{Kamihara-2008,Takahashi-2008}, Ba(Fe$_{1-x}$Co$_x$)$_2$As$_2$ \cite{Sefat-2008} or Sr$_{1-x}$La$_x$Fe$_2$As$_2$ \cite{Muraba-2010}. Recently the family of iron arsenide superconductors was enriched by the compounds \parent\ (1038, space group $P\overline{1}$) and polytypic  \zva\ (1048, space groups $P\overline{1}$, $P2_1/n$, $P4/n$) which have raised the chemical complexity  \cite{Loehnert-2011,Ni-2011,Kakiya-2011}. This new class recently expanded by analogous compounds with iridium (Ir1048) \cite{Kudo-2013} and palladium (Pd1038) \cite{Hieke-2013} instead of platinum. The crystal structures of 1038 and 1048 compounds are closely related, and can be rationalized as alternating stacking of iron arsenide and platinum arsenide layers separated by calcium ions as depicted in Fig.~\ref{fig:structure}. Platinum in the \ce{Pt3As8} or \ce{Pt4As8} layer, respectively, is nearly planar fourfold coordinated by arsenic, forming a twisted edge-sharing \ce{PtAs4} tile-pattern that contains As$_4^{2-}$-Zintl anions. The 1038 structure reveals one systematic Pt vacancy according to Pt$_3\square$As$_8$ (Fig.~\ref{fig:structure}) which is filled in the 1048 structure yielding \ce{Pt4As8} layers.

High critical temperatures up to 35 K were assigned to the 1048 compound, while Pt doping of the FeAs-layers in the 1038 phase induces superconductivity with critical temperatures below 15 K. Although superconductivity  induced by Pt-doping is known from other iron arsenides, the very different critical temperatures of the 1038 and 1048 phases were matter of discussion. Based on DFT calculations we suggested intrinsic electron doping of the 1048-phase \cite{Loehnert-2011} caused by the electrons coming from the additional Pt$^{2+}$ according to [($\overset{+2}{\rm{Ca}}\overset{+1.8}{\rm{Fe}}\overset{-3}{\rm{As}})_{10}]^{8+}$[($\overset{+2}{\rm{Pt}}_4\overset{-2}{\rm{As}}_8)]^{8-}$ (+0.2 $e^-$/Fe). In contrast, 1038 is a valence compound according to [($\overset{+2}{\rm{Ca}}\overset{+2}{\rm{Fe}}\overset{-3}{\rm{As}})_{10}]^{10+}$[($\overset{+2}{\rm{Pt}}_3\overset{-2}{\rm{As}}_8)]^{10-}$ and not superconducting. We have found that electron-doping of the 1038 phase is possible by lanthanum doping of the calcium site in (Ca$_{0.8}$La$_{0.2}$)$_{10}$(FeAs)$_{10}$(Pt$_3$As$_8$) with critical temperatures above 30 K \cite{Stuerzer-2012}.  The phase diagram of La-doped 1038 compounds is similar to those of the known iron arsenides \cite{Ni-2013}. Also the typical structural distortion of the 1038 parent compound \parent\ accompanied with magnetic ordering has recently been found \cite{Stuerzer-2013,Zhou-2013}. These results give clear evidence that electron doping of the FeAs layers in the 1038/1048 compounds can be realized either from the calcium layer or from the \ce{Pt4As8} layer. Hence, in spite of their structural complexity and low symmetry, these new superconductors turned out to be typical representatives of the iron arsenide family, with the 1038 phase as the magnetic non-superconducting parent compound \cite{Stuerzer-2013}.

A special structural feature arises in the 1038 structure due to the missing platinum in the Pt$_3\square$As$_8$ layer. The calcium site just above and below the platinum vacancy is eightfold anti-prismatically coordinated by arsenic, while the other calcium positions are surrounded by seven arsenic atoms only (see Fig.~\ref{fig:structure}). This particular calcium site reveals a distinct preference for $RE$ atoms, but with a remarkable size tolerance. These perquisites render the 1038-phase an ideal system to probe the rare earth dependent response of the compound to different sized substitutes, electron-doping, and strong magnetic impurities in \care\ with magnetic $RE^{3+}$ ions.

In this letter we show that superconductivity in $RE$ doped 1038 compounds can not only be induced by La-doping, \cite{Stuerzer-2012,Ni-2013} but also by the complete series of trivalent $RE$ ions in spite of their different size. The $T_c(x_{RE})$ phase diagrams of varying rare earth ions reveal a universal correlation of electron doping and critical temperature, but are independent of the type of $RE$ element. Furthermore the absence of superconductivity in \caeu\ containing Eu$^{2+}$ is in line with the above mentioned electron doping scenario. Finally the close similarity between 1048 and La-1038 are demonstrated by optimally doped La-1038 reaching the same $T_c$ than 1048.

\section{Experimental}

\begin{figure}[h]
\centering
\includegraphics[width=8cm]{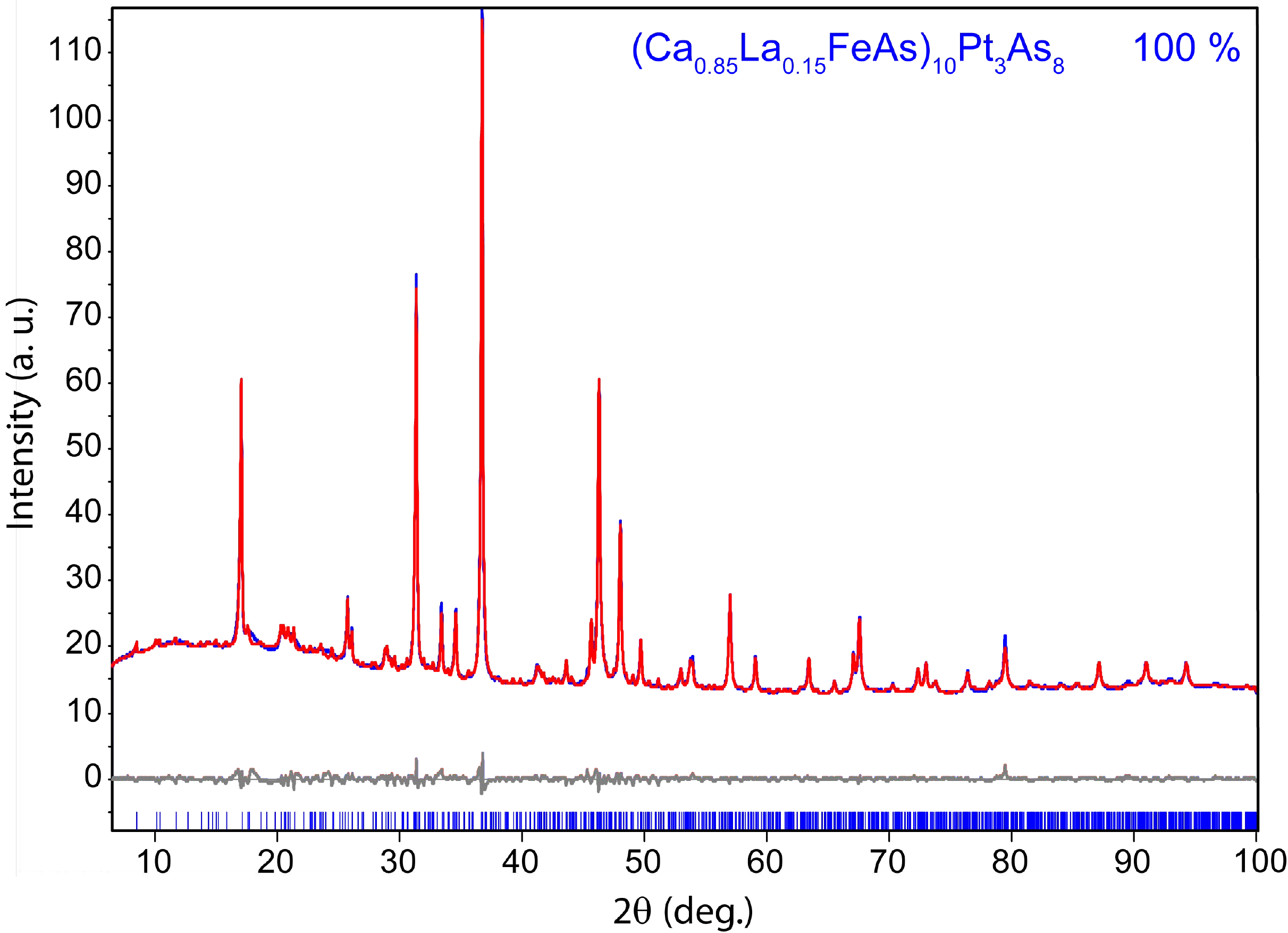}
\caption{\label{fig:xrd} X-ray powder pattern (blue), Rietveld fit (red), and difference curve (grey) of (Ca$_{0.85}$La$_{0.15}$)$_{10}$(FeAs)$_{10}$(Pt$_3$As$_8$)}
\end{figure}

Polycrystalline samples of rare earth calcium platinum iron-arsenides were synthesized as described in \cite{Stuerzer-2012}, and characterized by X-ray powder diffraction using the Rietveld method with TOPAS \cite{Topas}. Compositions were determined within errors of 10\% by refining occupation parameters and by X-ray spectroscopy (EDX). The X-ray powder pattern of (Ca$_{0.85}$La$_{0.15}$)$_{10}$(FeAs)$_{10}$(Pt$_3$As$_8$) together with the Rietveld fit is shown in Fig.~\ref{fig:xrd} as an example. Superconducting properties were determined using a $ac$-susceptometer at 1333 Hz in the temperature range of 3.5 to 300 K at a 3 Oe field. Magnetic susceptibility measurements of paramagnetic samples were performed on a Quantum Design MPMS XL5 SQUID magnetometer which allowed for measurements with fields up to 50 kOe at temperatures between 1.8 K and 300 K.\bigskip

\section{Results and Discussion}

\begin{figure}[h]
\centering
\includegraphics[width=8cm]{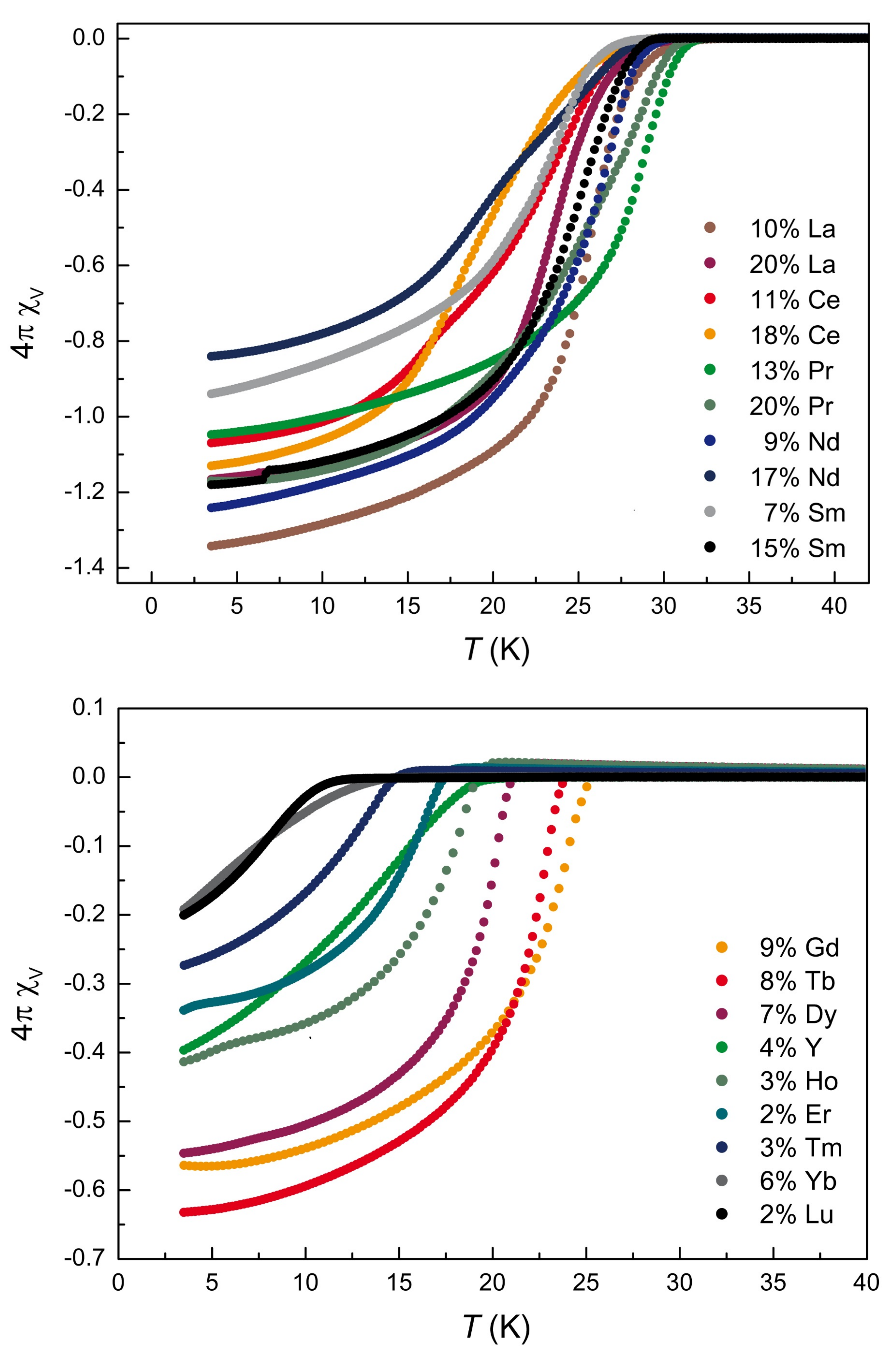}
\caption{\label{fig:mag-RE} ac susceptibility of the RE-1038 samples with $RE$ = La-Sm (top) and $RE$ = Gd-Tm (bottom)}
\end{figure}

\begin{figure}[h]
\centering
\includegraphics[width=8cm]{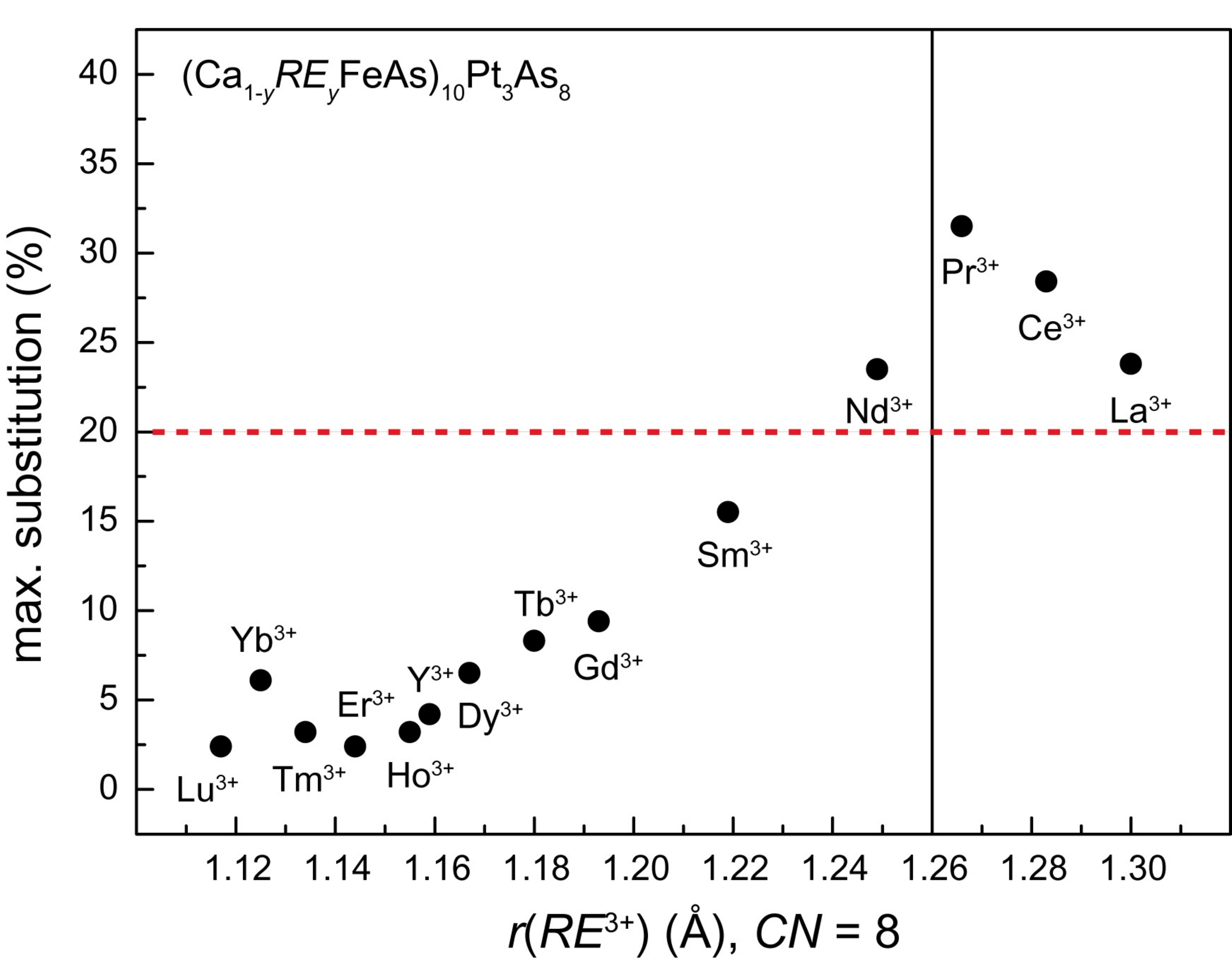}
\caption{\label{fig:solub-RE} Solubility of $RE$ dopants in \care\ against the ionic radius. The vertical line at 1.26 \AA\ marks the radius of Ca$^{2+}$.}
\end{figure}

Fig.~\ref{fig:mag-RE} shows the $ac$-susceptibility data for \care\ with $RE$ = Y, La-Sm, and Gd-Lu with nominal compositions $x$ = 0.1, 0.2 for the early, and $x$ = 0.2 for the late rare earth elements. Bulk superconductivity is detected in all samples. EDX measurements and Rietveld refinements confirmed the nominal composition of the early rare earth compounds La – Sm, whereas the structural tolerance towards rare earth substitution decreases with decreasing radii of the late rare earth elements Gd – Lu. Fig.~\ref{fig:solub-RE} depicts the maximum $RE$ solubility in \care. The red line marks the concentration level corresponding to a fully substituted eigthfold coordinated calcium position (20\%). The gradual decrease of the superconducting volume fractions (Fig.~\ref{fig:mag-RE}) for the late $RE$ elements (Gd-Lu) is caused by their limited solubility and increasing fractions of impurity phases.

Recent publications reported superconductivity in the La-doped compounds \cala\ with a maximum $T_c$ of 30 K \cite{Stuerzer-2012} and 26 K \cite{Ni-2013}, respectively. This finding was reasoned with a substitution of La$^{3+}$ preferably to the eightfold anti-prismatically coordinated Ca position in the structure, concomitant with electron doping to the FeAs-layer similar to La(O$_{1-x}$F$_x$)FeAs \cite{Stuerzer-2012}. A very similar doping scenario can be expected for the higher rare earth compounds, insofar the rare earth ions are trivalent.

Notably, the critical temperatures of the compounds containing La, Ce, Pr, Nd and Sm turned out to be  independent of the rare earth used, whereas $T_c$ drops significantly when the solubility limit restricts the $RE$ concentration for the late rare earth metals. Taking this into account, neither the kind of rare earth element used, nor its magnetic properties or effects on the structure perceptibly influence the superconducting properties, but exclusively its electronic contribution, \textit{i.e.} the electron transfer to the FeAs layers.

\begin{figure}[h]
\centering
\includegraphics[width=8cm]{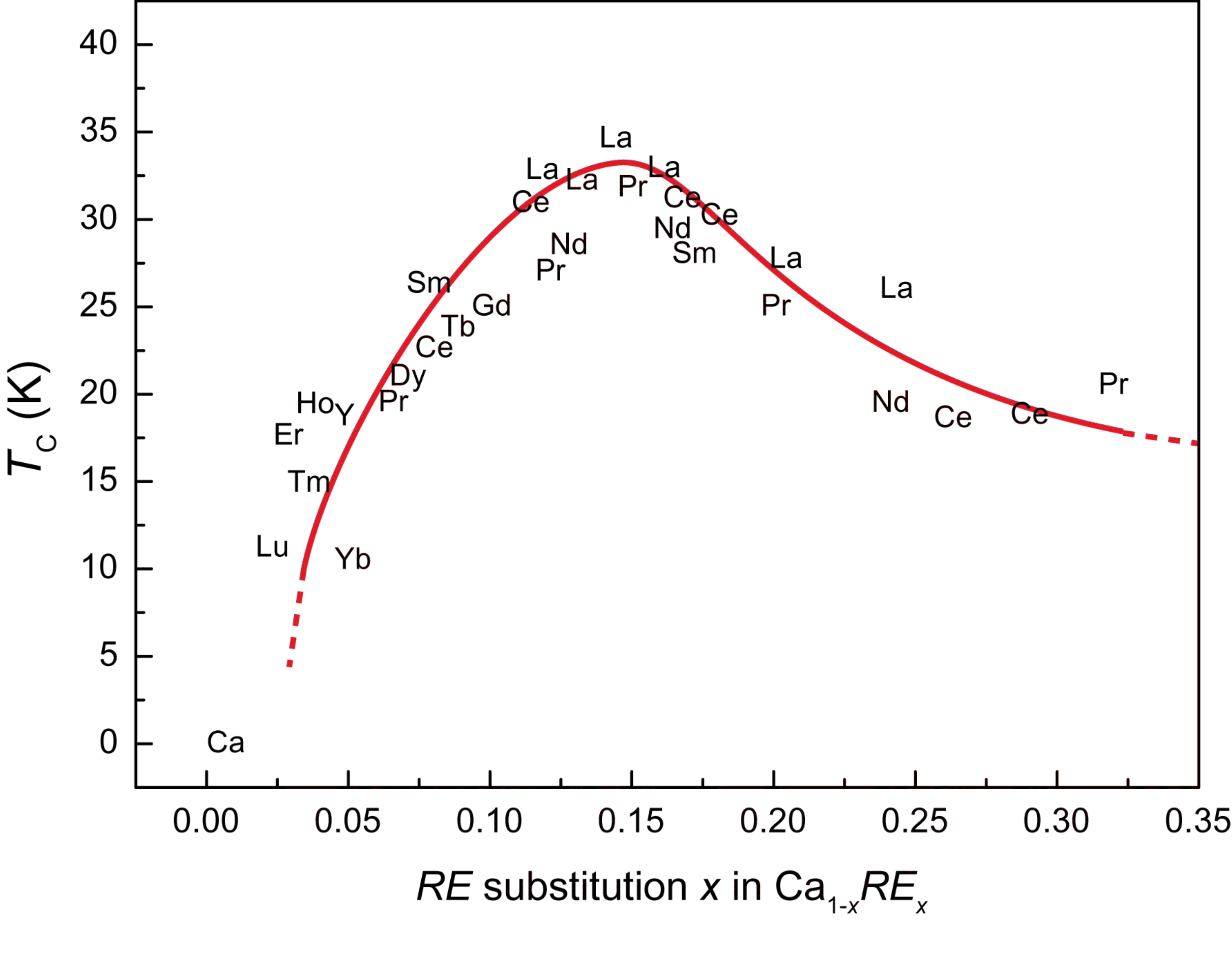}
\caption{\label{fig:RE-Tc-all} Critical temperatures of \care}
\end{figure}

The critical temperatures of the \care\ samples and the determined $RE$ concentrations are compiled in Fig.~\ref{fig:RE-Tc-all}. At $x$ = 0.13 a maximum of the critical temperature up to $T_c \approx$ 35 K is gained for La, Ce and Pr substituted compounds, whereas corresponding samples with heavier rare earth elements where not available so far. This consideration yields a substituent independent, universal curve rendering the influence of electron doping in the system \care\ and featuring a maximum of $T_c$ at $x$ = 0.13. Small deviations in $T_c$ may result from minor platinum substitution on the iron sites. The optimal electron doping level of 0.13$e^-$/FeAs is comparable to the 1111-superconductors La(O$_{1-x}$F$_x$)FeAs ($x$ = 0.11) \cite{Kamihara-2008} and Sm(O$_{1-x}$F$_x$)FeAs ($x$ = 0.1) \cite{Ren-2008}. The comparison with directly electron doped materials like Ba(Fe$_{1-x}$Co$_x$)$_2$As$_2$ appears not meaningful, since the extend of influence of chemical modification inside the FeAs-layer is still not fully understood.

\begin{figure}[h]
\begin{center}
\subfigure[]{\includegraphics[width=0.49\textwidth]{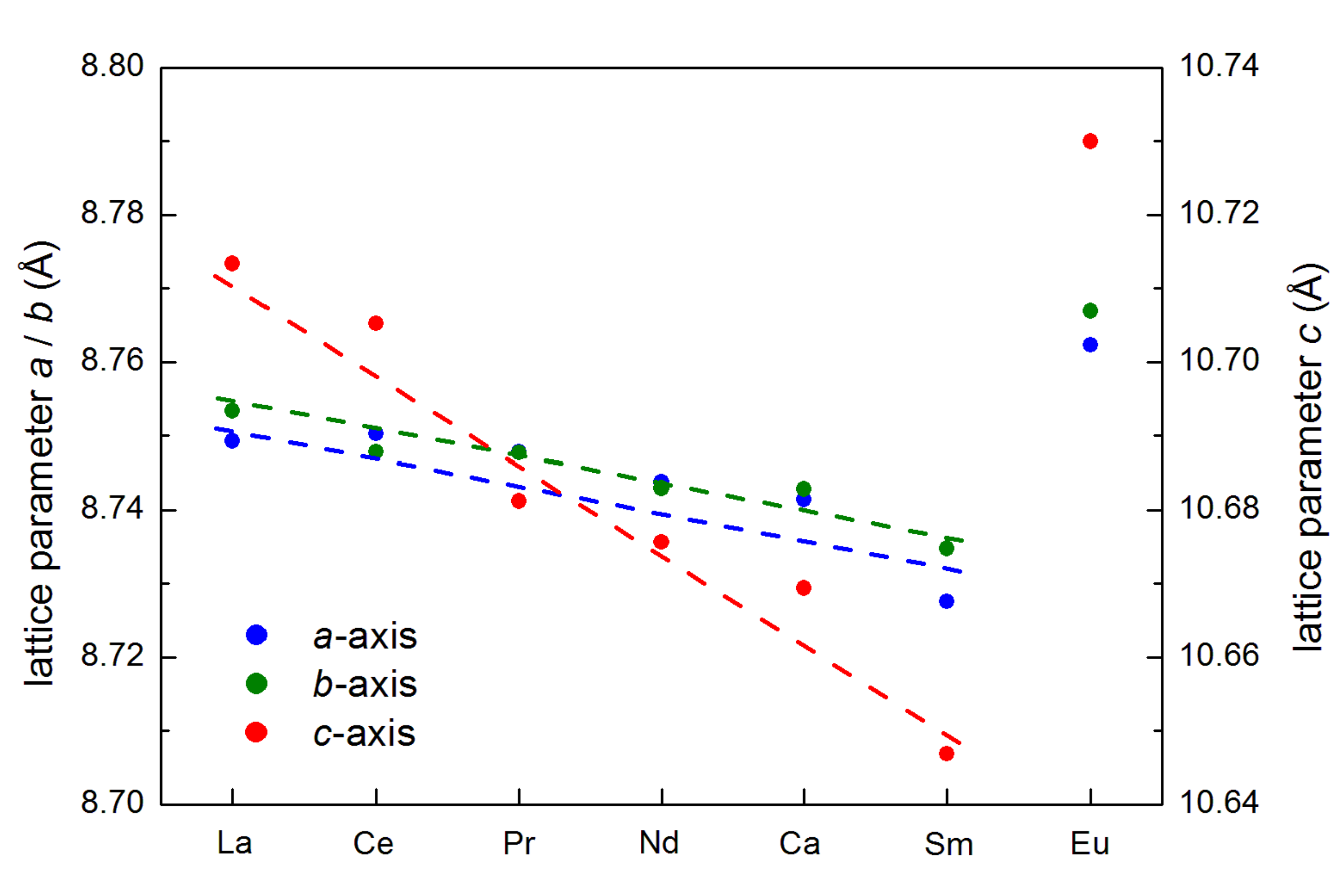}}
\subfigure[]{\includegraphics[width=0.49\textwidth]{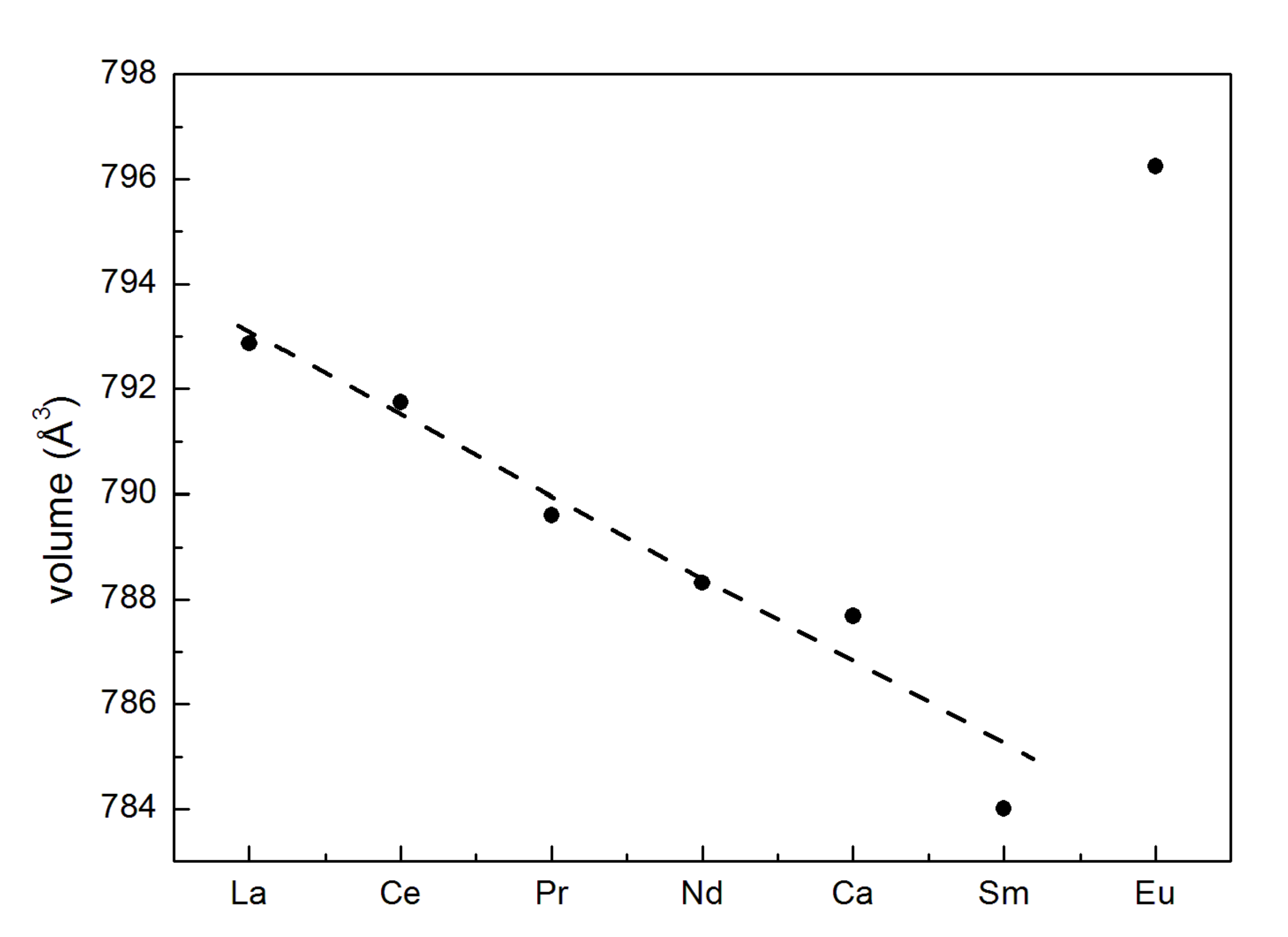}}
\caption{\label{fig:lattice} Lattice parameters (a) and cell volumes (b) of \care\ with $x$ = 0.2}
\end{center}
\end{figure}

It should be noted, that neglecting effects of size and magnetism of the rare earth elements on the superconducting properties appears oversimplified. One may expect at least an influence of strong magnetic moments present at the heavier elements like Holmium or Erbium affecting superconductivity. However, no suchlike was substantiated in our experiments. Even structural effects are minimal. Fig.~\ref{fig:lattice} depicts the evolution of the unit cell axes and volume for the compounds \care\ with $RE$ = La, Ce, Pr, Nd, Sm, Eu with constant rare earth concentration $x$ = 0.2. For comparison the values of the parent compound \parent\ were added at the position corresponding to the ionic radius of Ca$^{2+}$. Within the measurement accuracy the in plane parameters $a$ and $b$ remain equal and both decrease gradually with decreasing rare earth radii. In the same direction the contraction of the stacking axis $c$ increases more distinct, whereas all changes are well below 1\%.

\begin{figure}[h]
\centering
\includegraphics[width=8cm]{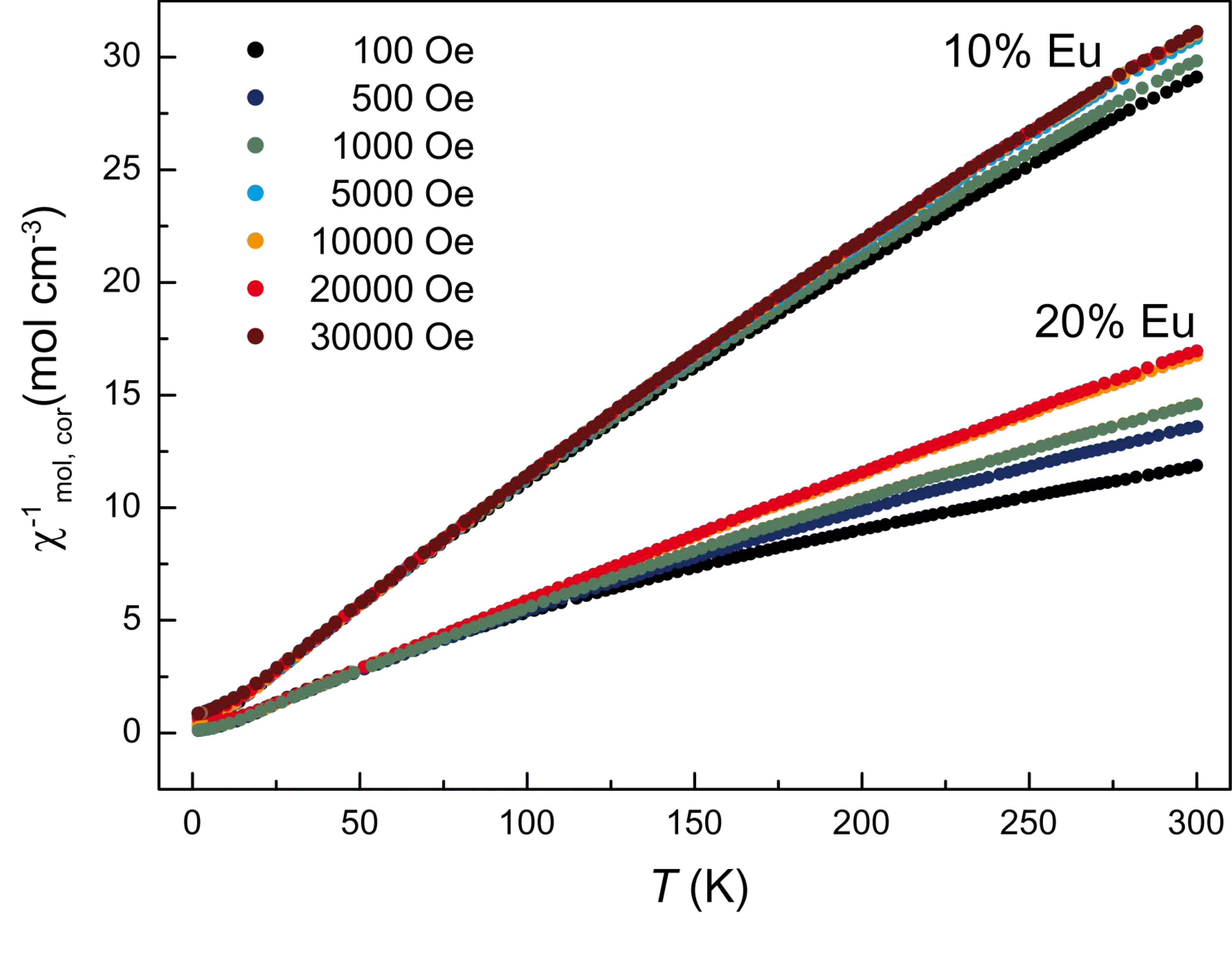}
\caption{\label{fig:Eu-mag} Magnetic susceptibilities of \caeu\ with $x$ = 0.1 and $x$ = 0.2}
\end{figure}

\begin{figure}[h]
\centering
\includegraphics[width=8cm]{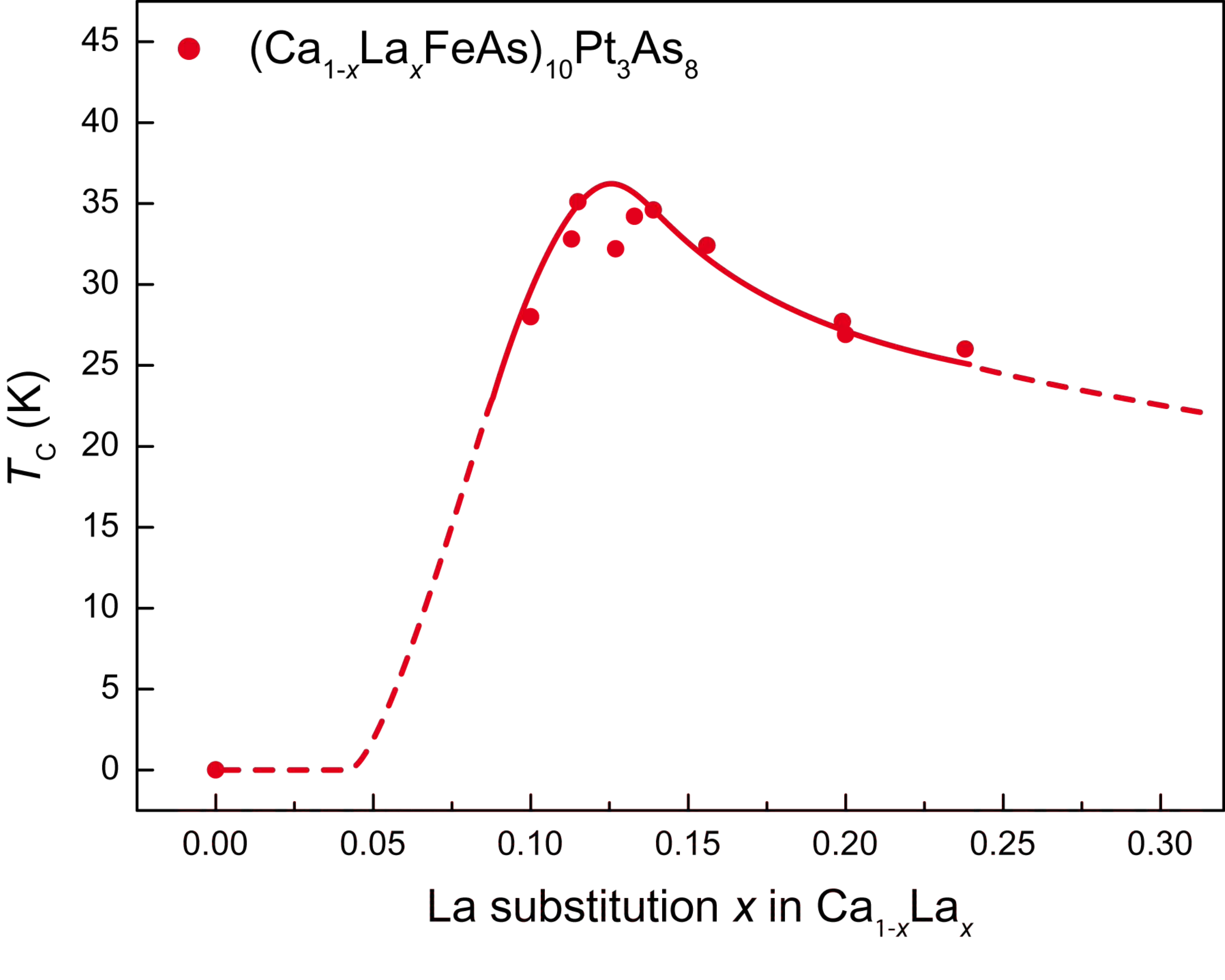}
\caption{\label{fig:La-phasedia} $T_c(x)$ phase diagram of \cala}
\end{figure}

The europium compounds \caeu\ are special cases of the $RE$ series due to the absence of superconductivity. The unit cell axis are significantly enlarged compared to the other $RE$ (Fig.~\ref{fig:lattice}), and magnetic measurements show paramagnetic behaviour. Fig.~\ref{fig:Eu-mag} displays the inverse magnetic susceptibilities of \caeu\ with $x$ = 0.1 and $x$ = 0.2  measured at different fields. Effective magnetic moments of 8.0 $\mu_{\rm B}$ per formula unit (Ca$_9$Eu)(FeAs)$_{10}$(Pt$_3$As$_8$) and (Ca$_8$Eu$_2$)(FeAs)$_{10}$(Pt$_3$As$_8$) were extracted from the Curie-Weiss fits, which are in excellent agreement with the effective moment of 7.94 $\mu_{\rm B}$ expected for Eu$^{2+}$. Thus our data are suggest divalent europium in line with the enhanced unit cell volume displayed in Fig.~\ref{fig:lattice}. Given the presence of Eu$^{2+}$, the absence of superconductivity in \caeu\ can clearly be referred to missing of electron doping of the FeAs-layer.


Fig.~\ref{fig:La-phasedia} depicts the $T_c(x)$ phase diagram of \cala\ with $x$ = 0.1 – 0.4. Analogous diagrams were obtained  with Ce and Pr doping (not shown). The preparation of homogeneous powder samples for $0 \leq x \leq 0.1$ were not successful so far due to phase separation into \parent\ and \cala\ with $x \geq 0.1$. A similar phase diagram was reported recently \cite{Ni-2013} based on single crystal data, identifying the same optimal doping level, whereas the maximum $T_c$ did not exceed 26 K due to considerable Pt mixing at the iron sites. \cala\ without Pt substitution features bulk superconductivity in the range investigated with a maximum $T_c$ of 35 K for $x \approx$ 0.13. Notably the critical temperature of 35 K coincides with $T_c$ reported for the 1048 compound \zva, as well as the electron doping level which was estimated by DFT calculation to be approximately 0.15~$e^-$/FeAs for the 1048 phase \cite{Stuerzer-2012}. This finding is fully consistent with the two different electron doping scenarios we suggested earlier, and emphasizes the close electronic relation between \cala\ and \zva\ despite their different crystal structures.

\section{Conclusion}

In summary our results show that superconductivity in the 1038 phase does not only occur by substitution of lanthanum for calcium, but by the complete series of trivalent rare earth elements Y, La-Sm, and Gd-Lu with appropriate radius. Superconductivity arises in all compounds investigated and depends only on the $RE$ concentration, but not on the type of $RE$. This gives proof for the  electron donor function of the $RE$ substitution. Other influences like lattice parameters and $RE$ magnetism are minimal, and show no measurable effect on the superconducting critical temperatures. This finding yields an universal $T_c(x)$ phase diagram for \care\  with a maximum $T_c$ of 35 K at $x$ = 0.13 independent of the kind of rare earth. Thereby the optimal doping level of $x$ = 0.13$e^-$/FeAs nearly coincides with known indirectly electron doped iron arsenides La(O$_{1-x}$F$_x$)FeAs ($x$ = 0.11) \cite{Kamihara-2008} and Sm(O$_{1-x}$F$_x$)FeAs ($x$ = 0.1) \cite{Ren-2008}. The absence of superconductivity in \caeu\ is rationalized with divalent Eu$^{2+}$ present in the structure and the absence of electron doping. Finally the close resemblance of electron doped \care\ and the 1048-phase \zva\ was demonstrated by featuring same $T_c^{\rm max} \approx$ 35 K coinciding at the same optimal doping level.

\section{Acknowledgement}

This work was financially supported by the German Research Foundation (DFG) within the priority program SPP 1458.

\section*{References}

\bibliographystyle{elsart-num}


\end{document}